\documentclass[%
 aip,
 amsmath,amssymb,
 reprint,%
]{revtex4-1}

\usepackage{graphicx}
\usepackage{dcolumn}
\usepackage{bm}

\usepackage[utf8]{inputenc}
\usepackage[T1]{fontenc}
\usepackage{mathptmx}

\begin{document}

\preprint{AIP/123-QED}

\title{Magnetic field dependence of the nonlocal spin Seebeck effect in Pt/YIG/Pt systems at low temperatures}

\author{Koichi Oyanagi}
\email{k.oyanagi@imr.tohoku.ac.jp}
 \affiliation{ 
Institute for Materials Research, Tohoku University, Sendai 980-8577, Japan
}%
\author{Takashi Kikkawa}%
 \affiliation{ 
Institute for Materials Research, Tohoku University, Sendai 980-8577, Japan
}%
  \affiliation{ 
WPI Advanced Institute for Materials Research, Tohoku University, Sendai 980-8577, Japan
}%

\author{Eiji Saitoh}
 \affiliation{ 
Institute for Materials Research, Tohoku University, Sendai 980-8577, Japan
}%
  \affiliation{ 
WPI Advanced Institute for Materials Research, Tohoku University, Sendai 980-8577, Japan
}%
\affiliation{ 
Center for Spintronics Research Network, Tohoku University, Sendai 980-8577, Japan
}%
\affiliation{ 
Department of Applied Physics, University of Tokyo, Hongo, Tokyo 113- 8656, Japan
}%
\affiliation{ 
Advanced Science Research Center, Japan Atomic Energy Agency, Tokai 319-1195, Japan
}%

\date{\today}

\begin{abstract}

We report the nonlocal spin Seebeck effect (nlSSE) in a lateral configuration of Pt/$\mathrm{Y}_3\mathrm{Fe}_5\mathrm{O}_{12}$(YIG)/Pt systems as a function of the magnetic field $B$ (up to 10 T) at various temperatures $T$ ($3~\textrm{K} < T < 300~\textrm{K} $). 
The nlSSE voltage decreases with increasing $B$ in a linear regime with respect to the input power (the applied charge-current squared $I^2$). 
The reduction of the nlSSE becomes substantial when the Zeeman energy exceeds thermal energy at low temperatures, which can be interpreted as freeze-out of magnons relevant for the nlSSE. 
Furthermore, we found the non-linear power dependence of the nlSSE with increasing $I$ at low temperatures ($T < 20~\textrm{K} $), at which the $B$-induced signal reduction becomes less visible.
Our experimental results suggest that in the non-linear regime high-energy magnons are over populated than those expected from the thermal energy. 
We also estimate the magnon spin diffusion length as functions of $B$ and $T$.
\end{abstract}

\maketitle
Spin caloritronics\cite{bauer2012spin} is an emerging field to study the interconversion between spin and heat currents. 
The spin Seebeck effect (SSE) is one of the fundamental phenomena in this field, referring to the spin-current generation from a heat current.
The SSE is well studied in a longitudinal configuration\cite{uchida2010observation, uchida2014longitudinal}, which consists of a heavy metal(HM)/ferromagnet(FM) bilayer system, typically Pt/$\mathrm{Y}_3\mathrm{Fe}_5\mathrm{O}_{12}$(YIG) junction. When a thermal gradient is applied perpendicular to the interface, a magnon spin current is generated in FM and converted into a conduction-electron spin current in HM via the interfacial exchange interaction\cite{tserkovnyak2005nonlocal}, which is subsequently detected as a transverse electric voltage via the inverse
spin Hall effect (ISHE)\cite{azevedo2005dc, saitoh2006conversion}.
Recent studies of the longitudinal SSE (LSSE)\cite{rezende2014magnon, kikkawa2015critical, kehlberger2015length} suggest that magnon transport in FM plays a key role in SSEs.

Nonlocal experiment is a powerful tool to investigate the transport of spin currents in various magnetic insulators\cite{giles2015long, cornelissen2015long, goennenwein2015non,velez2016competing, cornelissen2016temperature, zhou2017lateral, cornelissen2017nonlocal, yuan2018experimental, lebrun2018tunable, oyanagi2018efficient, xing2019magnon}. 
Especially, when a spin current is excited via a thermal gradient, it is called the nonlocal SSE (nlSSE)\cite{giles2015long, cornelissen2015long}. 
A typical nonlocal device consists of two HM wires on top of a magnetic insulator, which are electrically separated with the distance $d$. 
In nlSSE measurements, one of the HM wires is used as a heater; the Joule heating of an applied charge current ($I$) drives magnon spin currents in the magnetic insulator. 
Some of the magnons reach the other HM wire and inject a spin current, which is converted into a voltage via the ISHE. By changing the injector-detector separation distance $d$, we can address the transport property of magnon spin currents. 

In this paper, we report the high magnetic field ($B$) dependence of the nlSSE in lateral Pt/YIG/Pt systems at various temperatures from $T =$ 300 K to 3 K and up to $|B|$ = 10 T. We observed that the nlSSE signal $V_{\mathrm{2\omega}}$ decreases with increasing $B$, but the feature turns out to depend on the amplitude of the applied $I$.
In a linear regime ($V_{\mathrm{2\omega}}\propto I^2$), substantial $B$-induced suppression of the nlSSE was observed below 10 K, which is consistent with the previous LSSE results in Pt/YIG \cite{kikkawa2015critical, kikkawa2016complete}. 
In a non-linear regime ($V_{\mathrm{2\omega}} \not\propto I^2$), however, the nlSSE signal remains almost unchanged under high $B$ at low $T$s. By measuring the $d$ dependence, we estimate the magnon diffusion length $\lambda$ as functions of $T$ and $B$. 
 
\begin{figure}
\includegraphics{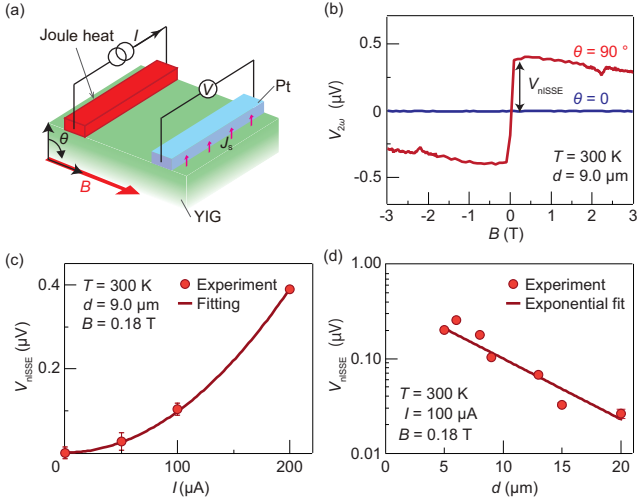}
\caption{\label{fig1} (a) A schematic illustration of the nlSSE measurement in a lateral Pt/YIG/Pt system. $B$, $\theta$, $I$, and $J_{\mathrm{s}}$ denote the external magnetic field, angle between $B$ and sample surface normal, charge current through the Pt injector, and spin current at the Pt/YIG detector interface, respectively.
An a.c. charge current is applied to the Pt injector, and the second harmonic voltage $V_{2\omega}$ is measured across the Pt detector. 
(b) The $B$ dependence of the nonlocal voltage $V_{2\omega}$ in the $d=9\ \mu$m sample. $V_{2\omega}$ at $\theta = 90^\circ$ ($\theta = 0$) is measured with $I = 200\ \mu$A ($100\ \mu$A). 
$V_{\mathrm{nlSSE}}= [V_{2\omega}(0.18\ \mathrm{T}) - V_{2\omega}(-0.18\ \mathrm{T})]/2$ represents the amplitude of the nlSSE. 
(c) $V_{\mathrm{nlSSE}}(I)$ in the $d=9\ \mu$m sample. The solid red line shows a $I^2$ fitting to data. 
(d) Semi logarithmic plot of $V_{\mathrm{nlSSE}}(d)$. The red line is fit with $V_{\mathrm{nlSSE}}= C \mathrm{exp}(-d/\lambda)$. The error bars represent the $68\%$ confidence level ($\pm$s.d.).}
\end{figure}

We prepared series of nonlocal Pt/YIG/Pt devices, schematically shown in Fig. \ref{fig1}(a). A 2.5-$\mu$m-thick YIG film was grown by liquid phase epitaxy on a $\mathrm{Gd}_3\mathrm{Ga}_5\mathrm{O}_{12}$ (111) substrate\cite{qiu2013spin}. 
On top of the YIG film, we fabricated two Pt wires using e-beam lithography and the lift-off process\cite{oyanagi2018efficient}. 
The dimension of the Pt wires is 200 $\mu$m length, 100 nm width, and 10 nm thickness. 
The Pt wires were deposited by magnetron sputtering in $\mathrm{Ar}^+$ atmosphere. 
We investigated four batches of samples (S1-S4) cut from the same YIG wafer. The $d$ dependence was studied in S1 ($d=5,\ 6,\ 8,\ 13,\ 15,$ and $20\ \mu$m) and S2 ($d=9\ \mu$m), while the $I$ dependence at low $T$s in S3 ($d=2\ \mu$m) and S4 ($d=8\ \mu$m). We measured  a nonlocal voltage using a lock-in detection technique; we applied an a.c. charge current, $I$, of 13.423 Hz in frequency to the injector Pt wire and measured a second harmonic nonlocal voltage $V_{2\omega}$ across the detector Pt wire\cite{cornelissen2015long}.

First, we confirmed that the obtained nonlocal voltage satisfies the features of the nlSSE at room temperature. Figure \ref{fig1}(b) shows typical $V_{2\omega}$ as a function of in-plane $B$ ($\theta = 90^\circ$) at 300 K in the $d=9\ \mu$m sample.
A clear $V_{2\omega}$ appears, whose sign changes with respect to the $B$ direction. $V_{2\omega}$ disappears when $B$ is applied perpendicular to the plane ($\theta = 0$). 
This symmetry is consistent with that of the SSE\cite{uchida2014longitudinal}. 
We define the low-field amplitude of the voltage signal as $V_{\mathrm{nlSSE}}= [V_{2\omega}(0.18\ \mathrm{T}) - V_{2\omega}(-0.18\ \mathrm{T})]/2$, at which the magnetization of YIG is fully saturated along $B$. 
As shown in Fig. \ref{fig1}(c), $V_{\mathrm{nlSSE}}$ is proportional to $I^2$, indicating that $V_{\mathrm{nlSSE}}$ appears due to the Joule heating. 
With increasing $B$, $V_{2\omega}$ gradually decreases, and at around $\pm2.2$ T, sharp dip structures show up, which are induced by magnon polarons due to magnon$-$TA-phonon hybridization\cite{kikkawa2016magnon, cornelissen2017nonlocal, shan2018enhanced}. 

By changing the injector-detector separation distance $d$, we estimate the length scale of the magnon spin current \cite{shan2017criteria}.  As shown in Fig. \ref{fig1}(d), $V_{\mathrm{nlSSE}}$ decreases with increasing $d$. A one-dimensional spin diffusion model\cite{cornelissen2015long, cornelissen2016magnon} describes the decay, which reads
\begin{equation}
V_{\mathrm{nlSSE}} = C\mathrm{exp}\left( -\frac{d}{\lambda} \right), \label{exp}
\end{equation}
where $\lambda$ is the magnon spin diffusion length and $C$ is the $d$-independent constant. 
We fit Eq. ($\ref{exp}$) to the $d$ dependence of $V_{\mathrm{nlSSE}}$ and obtain $\lambda = 6.76 \pm 0.16\ \mu$m at 300 K. Similar values are reported in previous studies in both thin (200 nm)\cite{cornelissen2015long}, and thick ($50\ \mu$m)\cite{shan2016influence} YIG films.

\begin{figure}
\includegraphics{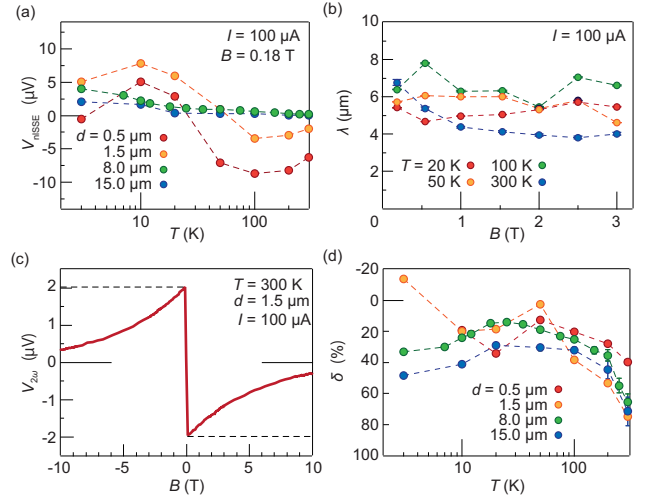}
\caption{\label{fig2} 
(a) Semi logarithmic plot of $V_{\mathrm{nlSSE}}(T)$ for various $d$ with $I=100\ \mu$A. 
(b) $\lambda(B)$ at various $T$s. We obtained $\lambda$ by $C \mathrm{exp}(-d/\lambda)$ fitting to the $d$ dependence of $V_{\mathrm{nlSSE}}$. (c) $V_{\mathrm{2\omega}}(B)$ in the $d=1.5\ \mu$m sample with $I= 100\ \mu$A at 300 K. (d) $\delta (T)$ for different $d$. $\delta$ is defined by Eq. (\ref{V}).}
\end{figure}

Next, we measured the $T$ dependence of $V_{2\omega}$ with $I=100\ \mu$A. As shown in Fig. \ref{fig2}(a), at 300 K negative voltages are observed for the $d=0.5$ and $1.5\ \mu$m samples, while the positive ones show up for the $d = 8$ and $15\ \mu$m samples.  
With decreasing $T$, the $d = 8$ and $15\ \mu$m samples exhibit a monotonic increase of $V_{\mathrm{nlSSE}}$. 
On the other hand, with decreasing $T$ the negative voltages observed for the $d=0.5$ and $1.5\ \mu$m samples at 300 K change their sign at  several tens Kelvin. 
The sign change of $V_\mathrm{{nlSSE}}$ with changing $d$ and $T$ has been observed in previous nlSSE experiments and explained as a result of a spatial profile of the magnon chemical potential $\mu_{\rm m}$ that governs the sign and amplitude of $V_\mathrm{{nlSSE}}$; a negative $\mu_{\rm m}$ created beneath the Pt injector exponentially decays apart from the injector and above a certain distance a positive one manifests due to the presence of YIG/GGG interface. The overall $\mu_{\rm m}$ profile varies with $T$\cite{ cornelissen2016magnon, shan2016influence, cornelissen2016temperature, zhou2017lateral, cornelissen2017nonlocal, shan2017criteria}. Furthermore, we found a second sign change for the $d= 0.5\ \mu$m sample at 3 K, which is unclear at this moment.

We now focus on the magnetic field $B$ dependent features of  $V_{2\omega}$. 
Figure \ref{fig2}(b) shows the $B$ dependence of $\lambda$ at various $T$s obtained by fitting Eq. ($\ref{exp}$) to $V_\mathrm{{nlSSE}}(d)$.
 At 300 K, $\lambda$ decreases with increasing $B$ by $30\ \%$ up to 3 T [from $\lambda=6.8\ \mu$m at $B = 0.18$ T to $4\ \mu$m at 3 T, see blue filled circles in Fig. \ref{fig2}(b)]. A similar field-induced decrease of $\lambda$ has been observed in the time-resolved LSSE\cite{hioki2017time}, nlSSE\cite{cornelissen2016magnetic}, and electrically excited magnon transport experiment\cite{cornelissen2016magnetic} at room temperature. 
On the other hand, at lower $T$s, $\lambda$ was found to be less sensitive to $B$ [see Fig. \ref{fig2}(b)]. 

\begin{figure*}
\includegraphics{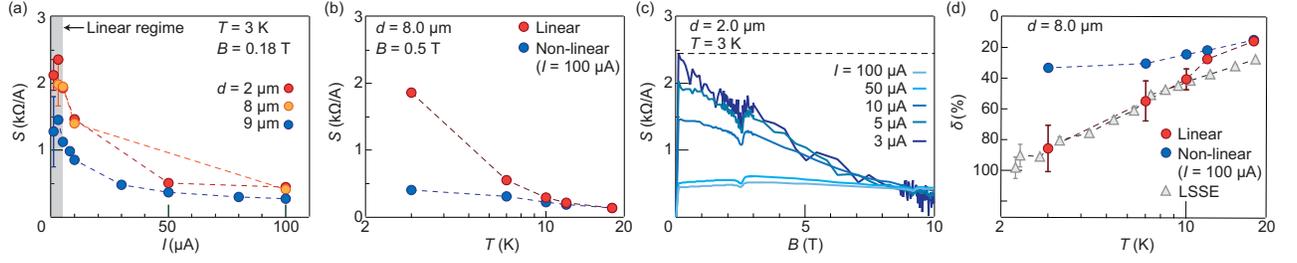}
\caption{\label{fig3}(a) $S$ at $B=0.18$ T and $T=3$ K for different $d$. $S$ is given as $S = V_{\mathrm{2\omega}}/I^2. \label{S}$ The gray shading represents the linear regime, where $S$ shows the linear dependence of $I^2$. (b) Semi logarithmic plot of $S$ in the linear (red circles) and non-linear (blue circles) regimes in the $d=8.0\ \mu$m sample at $B=0.5$ T. (c) $S$ with various $I$ in the $d=2.0\ \mu$m sample at $T=3$ K. (d) Semi logarithmic plot of $\delta (T)$ of the linear (red circles) and non-linear (blue circles) regimes in the $d=8\ \mu$m sample. The triangles are $\delta$ of the LSSE from Ref. \onlinecite{kikkawa2015critical}. The error bars represent the $68\%$ confidence level ($\pm$s.d.).}
\end{figure*}

To further investigate the effect of high $B$ on the nlSSE, we applied larger magnetic fields up to 10 T. 
Figure \ref{fig2}(c) shows a typical $V_{2\omega}-B$ result for $|B| < 10$ T in the $d = 1.5\ \mu$m sample with $I = 100\ \mu$A at 300 K.   
High $B$-induced suppression of $V_{\mathrm{2\omega}}$ is clearly observed. 
In Fig. \ref{fig2}(d) we plot the degree of $B$-induced $V_{2\omega}$ suppression up to 8 T, defined as 
\begin{equation}
\delta = 100 \times \left(1 - \frac{V_{\mathrm{2\omega}}^{8 \mathrm{T}}}{V_{\mathrm{2\omega}}^{0.18 \mathrm{T}}} \right) \label{V}
\end{equation}
as a function of $T$ for the $d=0.5$, $1.5$, $8.0$, and $15\ \mu$m samples. 
At 300 K, all the samples show the substantial high $B$-induced $V_{2\omega}$ reduction; $ 65\ \% < \delta < 75\ \%$ for the $d=1.5$, $8.0$, and $15\ \mu$m samples and $\delta= 39\ \%$ for the $d=0.5\ \mu$m sample. 
For the $d=8.0$ and $15\ \mu$m samples, with decreasing $T$, $\delta$ gradually decreases in the range of $20~\textrm{K} < T < 300~\textrm{K}$ and slightly increases below $20~\textrm{K}$. 
For the $d= 0.5$ and 1.5 $\mu$m samples, more complicated $T$ dependences were observed, which may be related to the non-monotonic $T$ responses of $V_{2\omega}$ as shown in Fig.  \ref{fig2}(a).  
The $T-\delta$ behavior above $20~\textrm{K}$ for the $d=8.0$ and $15\ \mu$m samples qualitatively agrees with the previous LSSE result in Pt/YIG-bulk systems \cite{kikkawa2015critical, jin2015effect}. 
However, below 20 K, the present nlSSE and previous LSSE results are totally different; $\delta$ of the LSSE becomes more outstanding with decreasing $T$ and reaches $\delta \sim 100 \%$ at $\sim 3~\textrm{K}$ \cite{kikkawa2016complete}, much greater than the present nlSSE results.

Significantly, we found that the disagreement at low temperatures is relevant to the applied current intensity $I$. So far, the nlSSE experiments were carried out with $I = 100\ \mu$A. Below 20 K, however, $V_{\mathrm{2\omega}}$ turned out to deviate from the $I^2$ scaling in this $I$ range. To see this, we introduce the normalization factor,
\begin{equation}
S= \frac{V_{\mathrm{2\omega}}}{I^2}. \label{S}
\end{equation}
If $V_{\mathrm{2\omega}}$ is proportional to $I^2$, $S$ keeps a constant with $I$, which was indeed confirmed above 20 K for $I < 100\ \mu$A. 
Figure \ref{fig3}(a) shows the $I$ dependence of $S$ at 3 K at the low $B$ of 0.18 T for the $d = 2,\ 8$, and $9\ \mu$m samples. $S$ takes almost the same value for $I\lesssim5\ \mu$A [see the gray colored area in Fig. \ref{fig3}(a)], but for $I \gtrsim5\ \mu$A, $S$ decreases with increasing $I$. 
We refer the former region to the linear regime ($V_{\mathrm{2\omega}} \propto I^2$), while the latter to the non-linear regime ($V_{\mathrm{2\omega}} \not\propto I^2$). 
In Fig. \ref{fig3}(b), we plot the $T$ dependence of $S$ in the linear and non-linear regimes at $B = 0.5~\textrm{T}$ for the $d=8\ \mu$m sample. The difference in $S$ between the linear and non-linear regimes becomes significant with decreasing $T$, and at 3 K $S$ in the linear regime is about 4 times greater than that in the non-linear regime. 
Importantly, the $B$ dependence of $V_{\mathrm{2\omega}}$ and $\delta$ also vary between the linear and non-linear regimes. 
In Fig. \ref{fig3}(c), we show representative results on $V_{\mathrm{2\omega}}$ versus $B$ with several $I$ values at 3 K for the $d=2.0\ \mu$m sample. 
In the linear regime (for $I=3\ \mu$A), clear $B$-induced $V_{\mathrm{2\omega}}$ suppression was observed ($\delta = 78\ \%$).
By increasing $I$ and entering into the non-linear regime, however, the $B$-induced $V_{\mathrm{2\omega}}$ reduction becomes less visible and, when $I=100\ \mu$A, $V_{\mathrm{2\omega}}$ is almost flat against $B$ ($\delta = -0.1\ \%$). 
In Fig. \ref{fig3}(d), we summarize the $\delta$ values as a function of $T$ obtained in the linear (red filled circles) and non-linear (blue filled circles) regimes for the $d = 8.0\ \mu$m sample and compare them to the previous LSSE result (gray filled triangles)\cite{kikkawa2016complete}. 
Interestingly, the $T$ dependence of $\delta$ for the nlSSE agrees well with that for the LSSE. 

The matching of the $T-\delta$ results in the low-$T$ range between the nlSSE in the linear regime and the LSSE indicates that the same mechanism governs the $B$-induced suppression. 
In Ref. \onlinecite{kikkawa2015critical, kikkawa2016complete}, the $T$ dependence of $\delta$ for the LSSE at low $T$s was well reproduced based on a conventional LSSE theory in which the effect of the Zeeman-gap opening in a magnon dispersion ($\propto g\mu_{\mathrm{B}}B$, where $g$ is the $g$-factor and $\mu_{\mathrm{B}}$ is the Bohr magneton) was taken into account;
the competition between thermal occupation of the magnon mode relevant for the LSSE (whose energy is of the order of $k_{\mathrm{B}}T$) and the Zeeman gap ($g\mu_{\mathrm{B}}B$) dominates the $B$-induced LSSE reduction. 
When $k_{\mathrm{B}}T \ll g\mu_{\mathrm{B}}B$ ($\approx 10 \ \mathrm{K}$ at 8 T), magnons cannot be thermally excited, leading to the suppression of the LSSE (see Fig. \ref{fig3}(d)). 
Our results indicate that the same scenario is valid also for the nlSSE in the linear regime.

Finally, we discuss the non-linear feature of the nlSSE. 
Both the $S$ and $\delta$ values of the nlSSE in the non-linear regime gradually increase with decreasing $T$ [see Figs. \ref{fig3}(b) and \ref{fig3}(d)]. However, their increasing rates are much smaller than those for the linear regime; both $S$ and $\delta$ at 3 K in the non-linear regime are $\sim 4$ times smaller than those at the same $T$ for the linear regime and also comparable to those at 12 K for the linear regime.  
These results suggest that the energy scale of magnons driving the nlSSE in the non-linear regime at 3 K may be much higher than the thermal energy $k_{\mathrm{B}}T$ at 3 K and the Zeeman energy $g\mu_{\mathrm{B}}B$ at 8 T. We note that, in the non-linear regime, the system temperature at least remains unchanged during the measurements, indicting that temperature rise due to the Joule heating is negligible. Furthermore, we found that, in the non-linear regime of $I=100\ \mu$A, the intensity of magnon-polaron dips at 3 K at $B =$ 2.5 T (9.2 T) is smaller (larger) than that in the linear regime of $I=3\ \mu$A at the same $T$ [see the dip structures marked by blue (red) triangles in Fig. \ref{fig3}(c)]. 
Here, the dip at the low $B$ (high $B$) originates from the spin currents carried by hybridized magnon$-$TA-phonon (magnon$-$LA-phonon) modes with the fixed energy of $E_{\rm MTA} \approx$ 6 K ($E_{\rm MLA} \approx$ 26 K).  
The dip intensity should thereby be maximized when the magnon mode at the energy of $E_{\rm MTA}$ ($E_{\rm MLA}$) is most significantly occupied under the condition of $k_{\mathrm{B}}T\approx$ 6 K (26 K), and apart from this temperature the intensity of magnon-polaron dip decreases. 
Therefore, the small (large) magnon-polaron dip at $B =$ 2.5 T (9.2 T) at 3 K in the non-linear regime also indicates the over occupation of high-energy magnons than that expected from the thermal energy $k_{\mathrm{B}}T$ at 3 K, as with the $S$ and $\delta$ results discussed above. 
Future work should address the origin of such high-nonequilibrium state realized in this regime. 

In summary, we systematically investigated the nonlocal spin Seebeck effect (nlSSE) in the lateral Pt/YIG/Pt systems as functions of separation distance ($d$), magnetic field ($B$), temperature ($T$), and excitation current ($I$). 
We found that below 20 K, the nlSSE voltage $V_{2\omega}$ deviates from the conventional $I^2$ scaling for $I\gtrsim5\ \mu$A. In this non-linear regime, the amplitudes of $V_{2\omega}$ and $B$-induced signal reduction $\delta$ become smaller than those in the linear regime, where $V_{2\omega} \propto I^2$ and $I<5\ \mu$A. 
In the linear regime, the $T$ dependence of $\delta$ of the nlSSE agrees well with that of the longitudinal SSE (LSSE), which can be attributed to the suppression of magnon excitation by the Zeeman effect. 
Our results provide an important clue in unraveling the $B$-induced suppression of the nlSSE and useful information on the non-linear effect in nonlocal spin transport at low temperatures.

We thank G. E. W. Bauer, B. J. van Wees, L. J. Cornelissen, J. Shan, T. Kuschel, F. Casanova, J. M. Gomez-Perez, S. Takahashi, Z. Qiu, Y. Chen, and R. Yahiro for fruitful discussion, and K. Nagase for technical help. This work is a part of the research program of ERATO Spin Quantum Rectification Project (No. JPMJER1402) from JST, the Grant-in-Aid for Scientific Research on Innovative Area Nano Spin Conversion Science (No. JP26103005), the Grant-in-Aid for Scientific Research (S) (No. JP19H05600), and Grant-in-Aid for Research Activity Start-up (No. JP19K21031) from JSPS KAKENHI, JSPS Core-to-Core program, the International Research Center for New-Concept Spintronics Devices, World Premier International Research Center Initiative (WPI) from MEXT, Japan. K.O. acknowledges support from GP-Spin at Tohoku University.

\nocite{*}
\bibliography{citation2}

\begin{thebibliography}{30}
\expandafter\ifx\csname natexlab\endcsname\relax\def\natexlab#1{#1}\fi
\expandafter\ifx\csname bibnamefont\endcsname\relax
  \def\bibnamefont#1{#1}\fi
\expandafter\ifx\csname bibfnamefont\endcsname\relax
  \def\bibfnamefont#1{#1}\fi
\expandafter\ifx\csname citenamefont\endcsname\relax
  \def\citenamefont#1{#1}\fi
\expandafter\ifx\csname url\endcsname\relax
  \def\url#1{\texttt{#1}}\fi
\expandafter\ifx\csname urlprefix\endcsname\relax\def\urlprefix{URL }\fi
\providecommand{\bibinfo}[2]{#2}
\providecommand{\eprint}[2][]{\url{#2}}

\bibitem[{\citenamefont{Bauer et~al.}(2012)\citenamefont{Bauer, Saitoh, and van
  Wees}}]{bauer2012spin}
\bibinfo{author}{\bibfnamefont{G.~E.~W.} \bibnamefont{Bauer}},
  \bibinfo{author}{\bibfnamefont{E.}~\bibnamefont{Saitoh}}, \bibnamefont{and}
  \bibinfo{author}{\bibfnamefont{B.~J.} \bibnamefont{van Wees}},
  \bibinfo{journal}{Nat. Mater.} \textbf{\bibinfo{volume}{11}},
  \bibinfo{pages}{391} (\bibinfo{year}{2012}).

\bibitem[{\citenamefont{Uchida et~al.}(2010)\citenamefont{Uchida, Adachi, Ota,
  Nakayama, Maekawa, and Saitoh}}]{uchida2010observation}
\bibinfo{author}{\bibfnamefont{K.}~\bibnamefont{Uchida}},
  \bibinfo{author}{\bibfnamefont{H.}~\bibnamefont{Adachi}},
  \bibinfo{author}{\bibfnamefont{T.}~\bibnamefont{Ota}},
  \bibinfo{author}{\bibfnamefont{H.}~\bibnamefont{Nakayama}},
  \bibinfo{author}{\bibfnamefont{S.}~\bibnamefont{Maekawa}}, \bibnamefont{and}
  \bibinfo{author}{\bibfnamefont{E.}~\bibnamefont{Saitoh}},
  \bibinfo{journal}{Appl. Phys. Lett.} \textbf{\bibinfo{volume}{97}},
  \bibinfo{pages}{172505} (\bibinfo{year}{2010}).

\bibitem[{\citenamefont{Uchida et~al.}(2014)\citenamefont{Uchida, Ishida,
  Kikkawa, Kirihara, Murakami, and Saitoh}}]{uchida2014longitudinal}
\bibinfo{author}{\bibfnamefont{K.}~\bibnamefont{Uchida}},
  \bibinfo{author}{\bibfnamefont{M.}~\bibnamefont{Ishida}},
  \bibinfo{author}{\bibfnamefont{T.}~\bibnamefont{Kikkawa}},
  \bibinfo{author}{\bibfnamefont{A.}~\bibnamefont{Kirihara}},
  \bibinfo{author}{\bibfnamefont{T.}~\bibnamefont{Murakami}}, \bibnamefont{and}
  \bibinfo{author}{\bibfnamefont{E.}~\bibnamefont{Saitoh}},
  \bibinfo{journal}{J. Phys. Condens. Matter} \textbf{\bibinfo{volume}{26}},
  \bibinfo{pages}{343202} (\bibinfo{year}{2014}).

\bibitem[{\citenamefont{Tserkovnyak et~al.}(2005)\citenamefont{Tserkovnyak,
  Brataas, Bauer, and Halperin}}]{tserkovnyak2005nonlocal}
\bibinfo{author}{\bibfnamefont{Y.}~\bibnamefont{Tserkovnyak}},
  \bibinfo{author}{\bibfnamefont{A.}~\bibnamefont{Brataas}},
  \bibinfo{author}{\bibfnamefont{G.~E.} \bibnamefont{Bauer}}, \bibnamefont{and}
  \bibinfo{author}{\bibfnamefont{B.~I.} \bibnamefont{Halperin}},
  \bibinfo{journal}{Rev. Mod. Phys.} \textbf{\bibinfo{volume}{77}},
  \bibinfo{pages}{1375} (\bibinfo{year}{2005}).

\bibitem[{\citenamefont{Azevedo et~al.}(2005)\citenamefont{Azevedo,
  Vilela~Le{\~a}o, Rodriguez-Suarez, Oliveira, and Rezende}}]{azevedo2005dc}
\bibinfo{author}{\bibfnamefont{A.}~\bibnamefont{Azevedo}},
  \bibinfo{author}{\bibfnamefont{L.~H.} \bibnamefont{Vilela~Le{\~a}o}},
  \bibinfo{author}{\bibfnamefont{R.~L.} \bibnamefont{Rodriguez-Suarez}},
  \bibinfo{author}{\bibfnamefont{A.~B.} \bibnamefont{Oliveira}},
  \bibnamefont{and} \bibinfo{author}{\bibfnamefont{S.~M.}
  \bibnamefont{Rezende}}, \bibinfo{journal}{J. Appl. Phys.}
  \textbf{\bibinfo{volume}{97}}, \bibinfo{pages}{10C715}
  (\bibinfo{year}{2005}).

\bibitem[{\citenamefont{Saitoh et~al.}(2006)\citenamefont{Saitoh, Ueda,
  Miyajima, and Tatara}}]{saitoh2006conversion}
\bibinfo{author}{\bibfnamefont{E.}~\bibnamefont{Saitoh}},
  \bibinfo{author}{\bibfnamefont{M.}~\bibnamefont{Ueda}},
  \bibinfo{author}{\bibfnamefont{H.}~\bibnamefont{Miyajima}}, \bibnamefont{and}
  \bibinfo{author}{\bibfnamefont{G.}~\bibnamefont{Tatara}},
  \bibinfo{journal}{Appl. Phys. Lett.} \textbf{\bibinfo{volume}{88}},
  \bibinfo{pages}{182509} (\bibinfo{year}{2006}).

\bibitem[{\citenamefont{Rezende et~al.}(2014)\citenamefont{Rezende,
  Rodr{\'\i}guez-Su{\'a}rez, Cunha, Rodrigues, Machado, Guerra, Ortiz, and
  Azevedo}}]{rezende2014magnon}
\bibinfo{author}{\bibfnamefont{S.~M.} \bibnamefont{Rezende}},
  \bibinfo{author}{\bibfnamefont{R.~L.}
  \bibnamefont{Rodr{\'\i}guez-Su{\'a}rez}},
  \bibinfo{author}{\bibfnamefont{R.~O.} \bibnamefont{Cunha}},
  \bibinfo{author}{\bibfnamefont{A.~R.} \bibnamefont{Rodrigues}},
  \bibinfo{author}{\bibfnamefont{F.~L.~A.} \bibnamefont{Machado}},
  \bibinfo{author}{\bibfnamefont{G.~A.~F.} \bibnamefont{Guerra}},
  \bibinfo{author}{\bibfnamefont{J.~C.~L.} \bibnamefont{Ortiz}},
  \bibnamefont{and} \bibinfo{author}{\bibfnamefont{A.}~\bibnamefont{Azevedo}},
  \bibinfo{journal}{Phys. Rev. B} \textbf{\bibinfo{volume}{89}},
  \bibinfo{pages}{014416} (\bibinfo{year}{2014}).

\bibitem[{\citenamefont{Kikkawa et~al.}(2015)\citenamefont{Kikkawa, Uchida,
  Daimon, Qiu, Shiomi, and Saitoh}}]{kikkawa2015critical}
\bibinfo{author}{\bibfnamefont{T.}~\bibnamefont{Kikkawa}},
  \bibinfo{author}{\bibfnamefont{K.}~\bibnamefont{Uchida}},
  \bibinfo{author}{\bibfnamefont{S.}~\bibnamefont{Daimon}},
  \bibinfo{author}{\bibfnamefont{Z.}~\bibnamefont{Qiu}},
  \bibinfo{author}{\bibfnamefont{Y.}~\bibnamefont{Shiomi}}, \bibnamefont{and}
  \bibinfo{author}{\bibfnamefont{E.}~\bibnamefont{Saitoh}},
  \bibinfo{journal}{Phys. Rev. B} \textbf{\bibinfo{volume}{92}},
  \bibinfo{pages}{064413} (\bibinfo{year}{2015}).

\bibitem[{\citenamefont{Kehlberger et~al.}(2015)\citenamefont{Kehlberger,
  Ritzmann, Hinzke, Guo, Cramer, Jakob, Onbasli, Kim, Ross, Jungfleisch
  et~al.}}]{kehlberger2015length}
\bibinfo{author}{\bibfnamefont{A.}~\bibnamefont{Kehlberger}},
  \bibinfo{author}{\bibfnamefont{U.}~\bibnamefont{Ritzmann}},
  \bibinfo{author}{\bibfnamefont{D.}~\bibnamefont{Hinzke}},
  \bibinfo{author}{\bibfnamefont{E.-J.} \bibnamefont{Guo}},
  \bibinfo{author}{\bibfnamefont{J.}~\bibnamefont{Cramer}},
  \bibinfo{author}{\bibfnamefont{G.}~\bibnamefont{Jakob}},
  \bibinfo{author}{\bibfnamefont{M.~C.} \bibnamefont{Onbasli}},
  \bibinfo{author}{\bibfnamefont{D.~H.} \bibnamefont{Kim}},
  \bibinfo{author}{\bibfnamefont{C.~A.} \bibnamefont{Ross}},
  \bibinfo{author}{\bibfnamefont{M.~B.} \bibnamefont{Jungfleisch}},
  \bibnamefont{et~al.}, \bibinfo{journal}{Phys. Rev. Lett.}
  \textbf{\bibinfo{volume}{115}}, \bibinfo{pages}{096602}
  (\bibinfo{year}{2015}).

\bibitem[{\citenamefont{Giles et~al.}(2015)\citenamefont{Giles, Yang, Jamison,
  and Myers}}]{giles2015long}
\bibinfo{author}{\bibfnamefont{B.~L.} \bibnamefont{Giles}},
  \bibinfo{author}{\bibfnamefont{Z.}~\bibnamefont{Yang}},
  \bibinfo{author}{\bibfnamefont{J.~S.} \bibnamefont{Jamison}},
  \bibnamefont{and} \bibinfo{author}{\bibfnamefont{R.~C.} \bibnamefont{Myers}},
  \bibinfo{journal}{Phys. Rev. B} \textbf{\bibinfo{volume}{92}},
  \bibinfo{pages}{224415} (\bibinfo{year}{2015}).

\bibitem[{\citenamefont{Cornelissen et~al.}(2015)\citenamefont{Cornelissen,
  Liu, Duine, Youssef, and van Wees}}]{cornelissen2015long}
\bibinfo{author}{\bibfnamefont{L.~J.} \bibnamefont{Cornelissen}},
  \bibinfo{author}{\bibfnamefont{J.}~\bibnamefont{Liu}},
  \bibinfo{author}{\bibfnamefont{R.~A.} \bibnamefont{Duine}},
  \bibinfo{author}{\bibfnamefont{J.~B.} \bibnamefont{Youssef}},
  \bibnamefont{and} \bibinfo{author}{\bibfnamefont{B.~J.} \bibnamefont{van
  Wees}}, \bibinfo{journal}{Nat. Phys.} \textbf{\bibinfo{volume}{11}},
  \bibinfo{pages}{1022} (\bibinfo{year}{2015}).

\bibitem[{\citenamefont{Goennenwein et~al.}(2015)\citenamefont{Goennenwein,
  Schlitz, Pernpeintner, Ganzhorn, Althammer, Gross, and
  Huebl}}]{goennenwein2015non}
\bibinfo{author}{\bibfnamefont{S.~T.} \bibnamefont{Goennenwein}},
  \bibinfo{author}{\bibfnamefont{R.}~\bibnamefont{Schlitz}},
  \bibinfo{author}{\bibfnamefont{M.}~\bibnamefont{Pernpeintner}},
  \bibinfo{author}{\bibfnamefont{K.}~\bibnamefont{Ganzhorn}},
  \bibinfo{author}{\bibfnamefont{M.}~\bibnamefont{Althammer}},
  \bibinfo{author}{\bibfnamefont{R.}~\bibnamefont{Gross}}, \bibnamefont{and}
  \bibinfo{author}{\bibfnamefont{H.}~\bibnamefont{Huebl}},
  \bibinfo{journal}{Applied Physics Letters} \textbf{\bibinfo{volume}{107}},
  \bibinfo{pages}{172405} (\bibinfo{year}{2015}).

\bibitem[{\citenamefont{V{\'e}lez et~al.}(2016)\citenamefont{V{\'e}lez,
  Bedoya-Pinto, Yan, Hueso, and Casanova}}]{velez2016competing}
\bibinfo{author}{\bibfnamefont{S.}~\bibnamefont{V{\'e}lez}},
  \bibinfo{author}{\bibfnamefont{A.}~\bibnamefont{Bedoya-Pinto}},
  \bibinfo{author}{\bibfnamefont{W.}~\bibnamefont{Yan}},
  \bibinfo{author}{\bibfnamefont{L.~E.} \bibnamefont{Hueso}}, \bibnamefont{and}
  \bibinfo{author}{\bibfnamefont{F.}~\bibnamefont{Casanova}},
  \bibinfo{journal}{Phys. Rev. B} \textbf{\bibinfo{volume}{94}},
  \bibinfo{pages}{174405} (\bibinfo{year}{2016}).

\bibitem[{\citenamefont{Cornelissen
  et~al.}(2016{\natexlab{a}})\citenamefont{Cornelissen, Shan, and van
  Wees}}]{cornelissen2016temperature}
\bibinfo{author}{\bibfnamefont{L.~J.} \bibnamefont{Cornelissen}},
  \bibinfo{author}{\bibfnamefont{J.}~\bibnamefont{Shan}}, \bibnamefont{and}
  \bibinfo{author}{\bibfnamefont{B.~J.} \bibnamefont{van Wees}},
  \bibinfo{journal}{Phys. Rev. B} \textbf{\bibinfo{volume}{94}},
  \bibinfo{pages}{180402} (\bibinfo{year}{2016}{\natexlab{a}}).

\bibitem[{\citenamefont{Zhou et~al.}(2017)\citenamefont{Zhou, Shi, Han, Yang,
  Rao, Zhang, Lang, Zhou, Pan, and Song}}]{zhou2017lateral}
\bibinfo{author}{\bibfnamefont{X.~J.} \bibnamefont{Zhou}},
  \bibinfo{author}{\bibfnamefont{G.~Y.} \bibnamefont{Shi}},
  \bibinfo{author}{\bibfnamefont{J.~H.} \bibnamefont{Han}},
  \bibinfo{author}{\bibfnamefont{Q.~H.} \bibnamefont{Yang}},
  \bibinfo{author}{\bibfnamefont{Y.~H.} \bibnamefont{Rao}},
  \bibinfo{author}{\bibfnamefont{H.~W.} \bibnamefont{Zhang}},
  \bibinfo{author}{\bibfnamefont{L.~L.} \bibnamefont{Lang}},
  \bibinfo{author}{\bibfnamefont{S.~M.} \bibnamefont{Zhou}},
  \bibinfo{author}{\bibfnamefont{F.}~\bibnamefont{Pan}}, \bibnamefont{and}
  \bibinfo{author}{\bibfnamefont{C.}~\bibnamefont{Song}},
  \bibinfo{journal}{Appl. Phys. Lett.} \textbf{\bibinfo{volume}{110}},
  \bibinfo{pages}{062407} (\bibinfo{year}{2017}).

\bibitem[{\citenamefont{Cornelissen et~al.}(2017)\citenamefont{Cornelissen,
  Oyanagi, Kikkawa, Qiu, Kuschel, Bauer, van Wees, and
  Saitoh}}]{cornelissen2017nonlocal}
\bibinfo{author}{\bibfnamefont{L.~J.} \bibnamefont{Cornelissen}},
  \bibinfo{author}{\bibfnamefont{K.}~\bibnamefont{Oyanagi}},
  \bibinfo{author}{\bibfnamefont{T.}~\bibnamefont{Kikkawa}},
  \bibinfo{author}{\bibfnamefont{Z.}~\bibnamefont{Qiu}},
  \bibinfo{author}{\bibfnamefont{T.}~\bibnamefont{Kuschel}},
  \bibinfo{author}{\bibfnamefont{G.~E.~W.} \bibnamefont{Bauer}},
  \bibinfo{author}{\bibfnamefont{B.~J.} \bibnamefont{van Wees}},
  \bibnamefont{and} \bibinfo{author}{\bibfnamefont{E.}~\bibnamefont{Saitoh}},
  \bibinfo{journal}{Phys. Rev. B} \textbf{\bibinfo{volume}{96}},
  \bibinfo{pages}{104441} (\bibinfo{year}{2017}).

\bibitem[{\citenamefont{Yuan et~al.}(2018)\citenamefont{Yuan, Zhu, Su, Yao,
  Xing, Chen, Ma, Lin, Shi, Shindou et~al.}}]{yuan2018experimental}
\bibinfo{author}{\bibfnamefont{W.}~\bibnamefont{Yuan}},
  \bibinfo{author}{\bibfnamefont{Q.}~\bibnamefont{Zhu}},
  \bibinfo{author}{\bibfnamefont{T.}~\bibnamefont{Su}},
  \bibinfo{author}{\bibfnamefont{Y.}~\bibnamefont{Yao}},
  \bibinfo{author}{\bibfnamefont{W.}~\bibnamefont{Xing}},
  \bibinfo{author}{\bibfnamefont{Y.}~\bibnamefont{Chen}},
  \bibinfo{author}{\bibfnamefont{Y.}~\bibnamefont{Ma}},
  \bibinfo{author}{\bibfnamefont{X.}~\bibnamefont{Lin}},
  \bibinfo{author}{\bibfnamefont{J.}~\bibnamefont{Shi}},
  \bibinfo{author}{\bibfnamefont{R.}~\bibnamefont{Shindou}},
  \bibnamefont{et~al.}, \bibinfo{journal}{Sci. Adv.}
  \textbf{\bibinfo{volume}{4}}, \bibinfo{pages}{eaat1098}
  (\bibinfo{year}{2018}).

\bibitem[{\citenamefont{Lebrun et~al.}(2018)\citenamefont{Lebrun, Ross, Bender,
  Qaiumzadeh, Baldrati, Cramer, Brataas, Duine, and
  Kl{\"a}ui}}]{lebrun2018tunable}
\bibinfo{author}{\bibfnamefont{R.}~\bibnamefont{Lebrun}},
  \bibinfo{author}{\bibfnamefont{A.}~\bibnamefont{Ross}},
  \bibinfo{author}{\bibfnamefont{S.~A.} \bibnamefont{Bender}},
  \bibinfo{author}{\bibfnamefont{A.}~\bibnamefont{Qaiumzadeh}},
  \bibinfo{author}{\bibfnamefont{L.}~\bibnamefont{Baldrati}},
  \bibinfo{author}{\bibfnamefont{J.}~\bibnamefont{Cramer}},
  \bibinfo{author}{\bibfnamefont{A.}~\bibnamefont{Brataas}},
  \bibinfo{author}{\bibfnamefont{R.~A.} \bibnamefont{Duine}}, \bibnamefont{and}
  \bibinfo{author}{\bibfnamefont{M.}~\bibnamefont{Kl{\"a}ui}},
  \bibinfo{journal}{Nature (London)} \textbf{\bibinfo{volume}{561}},
  \bibinfo{pages}{222} (\bibinfo{year}{2018}).

\bibitem[{\citenamefont{Oyanagi et~al.}(2018)\citenamefont{Oyanagi, Takahashi,
  Cornelissen, Shan, Daimon, Kikkawa, Bauer, van Wees, and
  Saitoh}}]{oyanagi2018efficient}
\bibinfo{author}{\bibfnamefont{K.}~\bibnamefont{Oyanagi}},
  \bibinfo{author}{\bibfnamefont{S.}~\bibnamefont{Takahashi}},
  \bibinfo{author}{\bibfnamefont{L.~J.} \bibnamefont{Cornelissen}},
  \bibinfo{author}{\bibfnamefont{J.}~\bibnamefont{Shan}},
  \bibinfo{author}{\bibfnamefont{S.}~\bibnamefont{Daimon}},
  \bibinfo{author}{\bibfnamefont{T.}~\bibnamefont{Kikkawa}},
  \bibinfo{author}{\bibfnamefont{G.~E.} \bibnamefont{Bauer}},
  \bibinfo{author}{\bibfnamefont{B.~J.} \bibnamefont{van Wees}},
  \bibnamefont{and} \bibinfo{author}{\bibfnamefont{E.}~\bibnamefont{Saitoh}},
  \bibinfo{journal}{arXiv preprint arXiv:1811.11972}  (\bibinfo{year}{2018}).

\bibitem[{\citenamefont{Xing et~al.}(2019)\citenamefont{Xing, Qiu, Wang, Yao,
  Ma, Cai, Jia, Xie, and Han}}]{xing2019magnon}
\bibinfo{author}{\bibfnamefont{W.}~\bibnamefont{Xing}},
  \bibinfo{author}{\bibfnamefont{L.}~\bibnamefont{Qiu}},
  \bibinfo{author}{\bibfnamefont{X.}~\bibnamefont{Wang}},
  \bibinfo{author}{\bibfnamefont{Y.}~\bibnamefont{Yao}},
  \bibinfo{author}{\bibfnamefont{Y.}~\bibnamefont{Ma}},
  \bibinfo{author}{\bibfnamefont{R.}~\bibnamefont{Cai}},
  \bibinfo{author}{\bibfnamefont{S.}~\bibnamefont{Jia}},
  \bibinfo{author}{\bibfnamefont{X.}~\bibnamefont{Xie}}, \bibnamefont{and}
  \bibinfo{author}{\bibfnamefont{W.}~\bibnamefont{Han}},
  \bibinfo{journal}{Phys. Rev. X} \textbf{\bibinfo{volume}{9}},
  \bibinfo{pages}{011026} (\bibinfo{year}{2019}).

\bibitem[{\citenamefont{Kikkawa
  et~al.}(2016{\natexlab{a}})\citenamefont{Kikkawa, Uchida, Daimon, and
  Saitoh}}]{kikkawa2016complete}
\bibinfo{author}{\bibfnamefont{T.}~\bibnamefont{Kikkawa}},
  \bibinfo{author}{\bibfnamefont{K.-i.} \bibnamefont{Uchida}},
  \bibinfo{author}{\bibfnamefont{S.}~\bibnamefont{Daimon}}, \bibnamefont{and}
  \bibinfo{author}{\bibfnamefont{E.}~\bibnamefont{Saitoh}},
  \bibinfo{journal}{J. Phys. Soc. Jpn.} \textbf{\bibinfo{volume}{85}},
  \bibinfo{pages}{065003} (\bibinfo{year}{2016}{\natexlab{a}}).

\bibitem[{\citenamefont{Qiu et~al.}(2013)\citenamefont{Qiu, Ando, Uchida,
  Kajiwara, Takahashi, Nakayama, An, Fujikawa, and Saitoh}}]{qiu2013spin}
\bibinfo{author}{\bibfnamefont{Z.}~\bibnamefont{Qiu}},
  \bibinfo{author}{\bibfnamefont{K.}~\bibnamefont{Ando}},
  \bibinfo{author}{\bibfnamefont{K.}~\bibnamefont{Uchida}},
  \bibinfo{author}{\bibfnamefont{Y.}~\bibnamefont{Kajiwara}},
  \bibinfo{author}{\bibfnamefont{R.}~\bibnamefont{Takahashi}},
  \bibinfo{author}{\bibfnamefont{H.}~\bibnamefont{Nakayama}},
  \bibinfo{author}{\bibfnamefont{T.}~\bibnamefont{An}},
  \bibinfo{author}{\bibfnamefont{Y.}~\bibnamefont{Fujikawa}}, \bibnamefont{and}
  \bibinfo{author}{\bibfnamefont{E.}~\bibnamefont{Saitoh}},
  \bibinfo{journal}{Appl. Phys. Lett.} \textbf{\bibinfo{volume}{103}},
  \bibinfo{pages}{092404} (\bibinfo{year}{2013}).

\bibitem[{\citenamefont{Kikkawa
  et~al.}(2016{\natexlab{b}})\citenamefont{Kikkawa, Shen, Flebus, Duine,
  Uchida, Qiu, Bauer, and Saitoh}}]{kikkawa2016magnon}
\bibinfo{author}{\bibfnamefont{T.}~\bibnamefont{Kikkawa}},
  \bibinfo{author}{\bibfnamefont{K.}~\bibnamefont{Shen}},
  \bibinfo{author}{\bibfnamefont{B.}~\bibnamefont{Flebus}},
  \bibinfo{author}{\bibfnamefont{R.~A.} \bibnamefont{Duine}},
  \bibinfo{author}{\bibfnamefont{K.}~\bibnamefont{Uchida}},
  \bibinfo{author}{\bibfnamefont{Z.}~\bibnamefont{Qiu}},
  \bibinfo{author}{\bibfnamefont{G.~E.~W.} \bibnamefont{Bauer}},
  \bibnamefont{and} \bibinfo{author}{\bibfnamefont{E.}~\bibnamefont{Saitoh}},
  \bibinfo{journal}{Phys. Rev. Lett.} \textbf{\bibinfo{volume}{117}},
  \bibinfo{pages}{207203} (\bibinfo{year}{2016}{\natexlab{b}}).

\bibitem[{\citenamefont{Shan et~al.}(2018)\citenamefont{Shan, Singh, Liang,
  Cornelissen, Galazka, Gupta, van Wees, and Kuschel}}]{shan2018enhanced}
\bibinfo{author}{\bibfnamefont{J.}~\bibnamefont{Shan}},
  \bibinfo{author}{\bibfnamefont{A.}~\bibnamefont{Singh}},
  \bibinfo{author}{\bibfnamefont{L.}~\bibnamefont{Liang}},
  \bibinfo{author}{\bibfnamefont{L.}~\bibnamefont{Cornelissen}},
  \bibinfo{author}{\bibfnamefont{Z.}~\bibnamefont{Galazka}},
  \bibinfo{author}{\bibfnamefont{A.}~\bibnamefont{Gupta}},
  \bibinfo{author}{\bibfnamefont{B.}~\bibnamefont{van Wees}}, \bibnamefont{and}
  \bibinfo{author}{\bibfnamefont{T.}~\bibnamefont{Kuschel}},
  \bibinfo{journal}{Appl. Phys. Lett.} \textbf{\bibinfo{volume}{113}},
  \bibinfo{pages}{162403} (\bibinfo{year}{2018}).

\bibitem[{\citenamefont{Shan et~al.}(2017)\citenamefont{Shan, Cornelissen, Liu,
  Youssef, Liang, and van Wees}}]{shan2017criteria}
\bibinfo{author}{\bibfnamefont{J.}~\bibnamefont{Shan}},
  \bibinfo{author}{\bibfnamefont{L.~J.} \bibnamefont{Cornelissen}},
  \bibinfo{author}{\bibfnamefont{J.}~\bibnamefont{Liu}},
  \bibinfo{author}{\bibfnamefont{J.~B.} \bibnamefont{Youssef}},
  \bibinfo{author}{\bibfnamefont{L.}~\bibnamefont{Liang}}, \bibnamefont{and}
  \bibinfo{author}{\bibfnamefont{B.~J.} \bibnamefont{van Wees}},
  \bibinfo{journal}{Phys. Rev. B} \textbf{\bibinfo{volume}{96}},
  \bibinfo{pages}{184427} (\bibinfo{year}{2017}).

\bibitem[{\citenamefont{Cornelissen
  et~al.}(2016{\natexlab{b}})\citenamefont{Cornelissen, Peters, Bauer, Duine,
  and van Wees}}]{cornelissen2016magnon}
\bibinfo{author}{\bibfnamefont{L.~J.} \bibnamefont{Cornelissen}},
  \bibinfo{author}{\bibfnamefont{K.~J.~H.} \bibnamefont{Peters}},
  \bibinfo{author}{\bibfnamefont{G.~E.~W.} \bibnamefont{Bauer}},
  \bibinfo{author}{\bibfnamefont{R.~A.} \bibnamefont{Duine}}, \bibnamefont{and}
  \bibinfo{author}{\bibfnamefont{B.~J.} \bibnamefont{van Wees}},
  \bibinfo{journal}{Phys. Rev. B} \textbf{\bibinfo{volume}{94}},
  \bibinfo{pages}{014412} (\bibinfo{year}{2016}{\natexlab{b}}).

\bibitem[{\citenamefont{Shan et~al.}(2016)\citenamefont{Shan, Cornelissen,
  Vlietstra, Youssef, Kuschel, Duine, and van Wees}}]{shan2016influence}
\bibinfo{author}{\bibfnamefont{J.}~\bibnamefont{Shan}},
  \bibinfo{author}{\bibfnamefont{L.~J.} \bibnamefont{Cornelissen}},
  \bibinfo{author}{\bibfnamefont{N.}~\bibnamefont{Vlietstra}},
  \bibinfo{author}{\bibfnamefont{J.~B.} \bibnamefont{Youssef}},
  \bibinfo{author}{\bibfnamefont{T.}~\bibnamefont{Kuschel}},
  \bibinfo{author}{\bibfnamefont{R.~A.} \bibnamefont{Duine}}, \bibnamefont{and}
  \bibinfo{author}{\bibfnamefont{B.~J.} \bibnamefont{van Wees}},
  \bibinfo{journal}{Phys. Rev. B} \textbf{\bibinfo{volume}{94}},
  \bibinfo{pages}{174437} (\bibinfo{year}{2016}).

\bibitem[{\citenamefont{Hioki et~al.}(2017)\citenamefont{Hioki, Iguchi, Qiu,
  Hou, Uchida, and Saitoh}}]{hioki2017time}
\bibinfo{author}{\bibfnamefont{T.}~\bibnamefont{Hioki}},
  \bibinfo{author}{\bibfnamefont{R.}~\bibnamefont{Iguchi}},
  \bibinfo{author}{\bibfnamefont{Z.}~\bibnamefont{Qiu}},
  \bibinfo{author}{\bibfnamefont{D.}~\bibnamefont{Hou}},
  \bibinfo{author}{\bibfnamefont{K.-i.} \bibnamefont{Uchida}},
  \bibnamefont{and} \bibinfo{author}{\bibfnamefont{E.}~\bibnamefont{Saitoh}},
  \bibinfo{journal}{Appl. Phys. Express} \textbf{\bibinfo{volume}{10}},
  \bibinfo{pages}{073002} (\bibinfo{year}{2017}).

\bibitem[{\citenamefont{Cornelissen and van
  Wees}(2016)}]{cornelissen2016magnetic}
\bibinfo{author}{\bibfnamefont{L.~J.} \bibnamefont{Cornelissen}}
  \bibnamefont{and} \bibinfo{author}{\bibfnamefont{B.~J.} \bibnamefont{van
  Wees}}, \bibinfo{journal}{Phys. Rev. B} \textbf{\bibinfo{volume}{93}},
  \bibinfo{pages}{020403} (\bibinfo{year}{2016}).

\bibitem[{\citenamefont{Jin et~al.}(2015)\citenamefont{Jin, Boona, Yang, Myers,
  and Heremans}}]{jin2015effect}
\bibinfo{author}{\bibfnamefont{H.}~\bibnamefont{Jin}},
  \bibinfo{author}{\bibfnamefont{S.~R.} \bibnamefont{Boona}},
  \bibinfo{author}{\bibfnamefont{Z.}~\bibnamefont{Yang}},
  \bibinfo{author}{\bibfnamefont{R.~C.} \bibnamefont{Myers}}, \bibnamefont{and}
  \bibinfo{author}{\bibfnamefont{J.~P.} \bibnamefont{Heremans}},
  \bibinfo{journal}{Phys. Rev. B} \textbf{\bibinfo{volume}{92}},
  \bibinfo{pages}{054436} (\bibinfo{year}{2015}).

\end{thebibliography}
\end{document}